\documentclass[usenatbib]{mnras}
\usepackage{graphicx}
\usepackage{natbib}
\usepackage{latexsym}
\usepackage{amsmath}
\usepackage{setspace}
\usepackage{booktabs}
\usepackage{array, longtable}
\usepackage[bf, singlelinecheck=off]{caption}
\usepackage{subfig}
\usepackage[T1]{fontenc}
\usepackage{aecompl}
\usepackage{hyperref}
\usepackage{gensymb}

\newcommand{\etal}{et~al.} 
\newcommand{\ionhy}{H{\sc ii} }

\newcommand{\kms}{$\mbox{km~s}^{-1}$}

\newcommand{\specsfig}[1]        
{
   \begin{center}
     \begin{minipage}[t]{0.45\textwidth}
         \psfig{file=#1.eps,height=0.9\textwidth,angle=270}
     \end{minipage}
     \end{center}
 }

\newcommand{\specdfig}[2]        
{
   \begin{center}
     \begin{minipage}[t]{0.45\textwidth}
         \psfig{file=#1.eps,height=0.9\textwidth,angle=270}
     \end{minipage}
     \hfill
     \begin{minipage}[t]{0.45\textwidth}
         \psfig{file=#2.eps,height=0.9\textwidth,angle=270}
     \end{minipage}
   \end{center}
}

\begin{document}
\title[Molecular line search toward flaring 6.7-GHz methanol masers]{Molecular line search toward the flaring 6.7-GHz methanol masers of  G\,24.33+0.13  and G\,359.62--0.24: rare maser transitions detected}
\author[McCarthy \etal]{
	T.\ P. McCarthy,$^{1}$\thanks{Email: tiegem@utas.edu.au}
	G. Orosz,$^{1}$,
	S.\ P. Ellingsen,$^{1}$
	S.\ L. Breen,$^{2}$
	M.\ A. Voronkov,$^{3}$
	R.\ A. Burns,$^{4,5}$
	\newauthor M. Olech,$^{6}$
	Y. Yonekura,$^{7}$ 
	T. Hirota,$^{4,8}$
	L.\ J. Hyland,$^{1}$
	and P. Wolak$^{9}$
  \\
  \\
  $^1$ School of Natural Sciences, University of Tasmania, Private Bag 37, Hobart, Tasmania 7001, Australia\\
  $^2$ SKAO, Jodrell Bank, Lower Withington, Macclesfield,
Cheshire, SK11 9FT, UK\\
  $^3$ CSIRO Space \& Astronomy, PO Box 76, Epping, NSW 2121, Australia\\
  $^{4}$ Mizusawa VLBI Observatory, National Astronomical Observatory of Japan, 2-21-1 Osawa, Mitaka, Tokyo 181-8588, Japan \\
  $^{5}$ Korea Astronomy and Space Science Institute, 776 Daedeokdae-ro, Yuseong-gu, Daejeon 34055, Republic of Korea\\
  $^{6}$ Space Radio-Diagnostics Research Centre,  University of Warmia and Mazury, ul.Oczapowskiego 2, 10-719 Olsztyn, Poland \\
  $^{7}$ Center for Astronomy, Ibaraki University, 2-1-1 Bunkyo, Mito, Ibaraki 310-8512, Japan \\
  $^{8}$ Department of Astronomical Sciences, SOKENDAI (The Graduate University for Advanced Studies), Osawa 2-21-1, Mitaka-shi, Tokyo 181-8588, Japan \\
  $^{9}$ Institute of Astronomy, Faculty of Physics, Astronomy and Informatics, Nicolaus Copernicus University, Grudziadzka 5, 87-100 Torun, Poland\\}

 \maketitle

\begin{abstract}

We have performed a molecular line search toward the flaring 6.7-GHz masers G\,24.33+0.13 and G\,359.62--0.24 using the Australia Telescope Compact Array. We present spectra of the 6.7-GHz class~II methanol and 22.2-GHz water masers toward these sources and provide comparison with other recent flaring events these sources have experienced. We also detect the fourth example of a 23.4-GHz class~I methanol maser, and the eleventh example of a 4.8-GHz formaldehyde maser toward G\,24.33+0.13. Alongside these results, we observe the previously detected ammonia (3,3) emission and report upper limits on the presence of various other cm-wavelength methanol, ammonia and OH transitions. Our results are consistent with the flaring of G\,24.33+0.13 being driven by a variable accretion rate in the host high-mass young stellar object.

\end{abstract}


\section{Introduction}

Astronomical maser emission has been detected from a wide array of different transitions from the methanol molecular species \citep[e.g.][]{Barrett+71,Wilson+84,Wilson+85, Morimoto+85,Menten91b,Breen+19}. These masers have proven useful tools for investigating the evolutionary phases and underlying conditions of star-formation regions within our Galaxy \citep[e.g.][]{Fontani+10, Ellingsen+11a, Breen+13b, Voronkov+14, Leurini+16, McCarthy+18}. Methanol masers are pumped through either collisional or radiative interactions, which  divides them into class~I and class~II respectively \citep{Batrla+87, Menten91b, Cragg+92, Cragg+05, Sobolev+97a,  Leurini+16}.

The 6.7-GHz class~II transition is useful for the study of high-mass star-formation regions as they are typically highly luminous, and exclusively associated with high-mass star formation \citep{Breen+13b}. The class~II masers form close to the young stellar objects (YSOs) where they are radiatively pumped via infrared emission from the protostars \citep{Walsh+98, Cragg+05}. 
Variability in the class~II methanol maser line is relatively common on timescales of months to years \citep{Caswell+95d, Ellingsen07}. However, these variations are generally not extreme, with peak intensities remaining similar over their timescales of 20 years \citep{Ellingsen07}.  \citet{Goedhart+04} reported the detection of periodic variability in 7 class~II maser sources, along with the first detection of rapid and extreme flux density increase (referred to as flares) from the 6.7-GHz line toward several sources (e.g. G\,337.92-0.46, G\,345.00-0.22, G\,359.61-0.24). Although still a rare phenomenon, over the past two decades several flares have been  detected from the 6.7-GHz maser line \citep[e.g. ][]{Fujisawa+15, Macleod+18, Sugiyama+19}. The transient nature of these flares makes it difficult to determine the exact mechanisms responsible for the events. There are three commonly suggested explanations of this phenomenon: increase in continuum source photons; increases in path length of maser cloud along line of sight or underlying maser pumping conditions improving \citep{Deguchi&89, Burns+20a, Burns+20b}.

A link between protostellar accretion outbursts and maser flares has been made in three class~II methanol maser sources to date: S255 NIRS 3 \citep{Moscadelli+17}, NGC~6334I-MM1  \citep{Hunter+18} and G\,358.93--0.03 \citep{Burns+20b}. These events in protostars, where accretion rates burst higher than average, are thought to be necessary for the production of high-mass stars \citep{Meyer+17}. These events result in an enhancement of the radiation field which drives the increased luminosity of the class~II lines through induction of more favourable pumping conditions. The maser flare associated with the accretion burst in G\,358.93--0.03 is of particular importance, with an extensive maser line survey during the flaring period providing the first astronomical detections of 6 class~II methanol maser lines, including the first ever detections of torsionally excited maser lines \citep{Breen+19, Brogan+19, Chen+20}.

In this paper we report the results of molecular line searches (between $\sim6-10$~GHz and $\sim19.9-24.2$~GHz) towards two recent 6.7-GHz maser flaring events associated with high-mass star-formation (HMSF) regions G\,24.33+0.13 and G\,359.62--0.24. Both of these sources have had methanol, water and OH masers previously detected \citep[e.g. ][]{Caswell+95c,Caswell98, Szymczak+05, Breen+10b, Chen+11}. Flaring in the class~II masers of G\,24.33+0.13 has been previously detected in December 2010 by \citet{Wolak+18}, in an event that lasted for approximately 450 days. The flare event in G\,24.33+0.13 that we report here was first identified by \citet{Wolak+19} on the 5th of September 2019. Monitoring by \citet{Goedhart+04} over the period from 1999 January to 2003 March detected flaring in a single maser component toward G\,359.62--0.24, starting in 2000 January and lasting for $\sim500$ days, with the other maser components of this source not varying significantly over this time. The current flaring event in G\,359.62--0.24 was first reported by the Ibaraki 6.7-GHz Methanol Maser Monitor (iMet) program on the 13th of July 2020 \citep{Yonekura+16}. Notifications of both flaring events were received as part of the collaboration in the M2O program.\footnote{The Maser Monitoring Organisation (M2O) is a global cooperative of maser monitoring programs. See MaserMonitoring.org.}

\section{Observations} \label{sec:observations}

\begin{table*}
	\begin{center}
		\caption{Details of the observation epochs. Right ascension and declination coordinates represent the adopted target position for our observations. Synthesised beam sizes listed with respect to the 6.7-GHz class~II methanol maser transition for the $4-10$~GHz setup, and the 22.2-GHz water maser line for the $19-24$~GHz setup.}
		\begin{tabular}{lllllccc} \hline
			\multicolumn{1}{c}{\bf Source}& \multicolumn{1}{c}{\bf Epoch} & \multicolumn{1}{c}{\bf Array} &  \multicolumn{1}{c}{\bf R.A.}&  \multicolumn{1}{c}{\bf Dec.}&  \multicolumn{1}{c}{\bf Synthesised} &\multicolumn{1}{c}{\bf Total Obs.} &\multicolumn{1}{c}{\bf On-source}  \\ 
			\multicolumn{1}{c}{\bf } & \multicolumn{1}{c}{}& \multicolumn{1}{c}{\bf Config.} & \multicolumn{1}{c}{\bf $h$~~~$m$~~~$s$}& \multicolumn{1}{c}{\bf $^\circ$~~~$\prime$~~~$\prime\prime$} &  \multicolumn{1}{c}{\bf Beam Size} & \multicolumn{1}{c}{\bf Length} & \multicolumn{1}{c}{\bf Time}\\  \hline
			\multicolumn{8}{c}{$4-10$~GHz} \\\hline 
			G\,24.33+0.13 & 2019 Nov & 1.5C & 18 35 08.1 & --07 35 03.6 &  $ 84.2^{\prime\prime}\times4.0'', 1.5\degree$ & 5.5 hrs & 174 min \\
			& 2021 Feb & 6D & 18 35 08.1 & --07 35 03.6 &  $79.6''\times2.5'', 6.6\degree$ & 3 hrs & 109 min \\ 
			 G\,359.62--0.24 & 2020 Jul & H214  & 17 45 39.1 & --29 23 30.0 & $35.0''\times20.7'', 64\degree$ & 6 hrs & 250 min\\ \hline
			 \multicolumn{8}{c}{$19-24$~GHz} \\\hline
			 G\,24.33+0.13 & 2019 Nov & 1.5C & 18 35 08.1 & --07 35 03.6 &  $63.7''\times2.2'',4.0\degree$ & 6.6 hrs & 165 min  \\ \hline
			 
		\end{tabular} \label{tab:array_config}		
	\end{center}
\end{table*}

\begin{table*}
	\begin{center}
		\caption{Details of the observed transitions towards G\,24.33+0.13 and G\,359.62--0.24 along with velocity range and RMS noise of the final spectral line cubes. Detected transitions have their details marked in bold. Multiple spectral resolutions were used when searching for the presence of emission, however, as a reference we provide the RMS noise values for all lines using a spectral resolution of 0.1~\kms\ (for the 4--10~GHz setup) or 0.5~\kms\ (for the higher frequency setup; denoted by a `$^\dagger$'). Uncertainty in the last digit of the adopted rest frequency is indicated by the value in parentheses. Note that the velocity range of the 7.7-GHz spectral line cube for G\,24.33+0.13 does not cover the range where emission is observed from other class~II methanol lines in this source. This is due to an error in the CABB frequency setup for these observations. Hyphens represent transitions that were not observed in a particular source.}
		\begin{tabular}{lllcccc} \hline
			\multicolumn{1}{c}{\bf Molecular}& \multicolumn{1}{c}{\bf Transition} & \multicolumn{1}{c}{\bf Rest Frequency}  &\multicolumn{2}{c}{\bf G\,24.33+0.13}&\multicolumn{2}{c}{\bf G\,359.62--0.24} \\
			\multicolumn{1}{c}{\bf Species} & \multicolumn{1}{c}{} & \multicolumn{1}{c}{} & \multicolumn{1}{c}{\bf Vel. Range}& \multicolumn{1}{c}{\bf RMS Noise}& \multicolumn{1}{c}{\bf Vel. Range}& \multicolumn{1}{c}{\bf RMS Noise} \\ 
			\multicolumn{1}{c}{\bf } & \multicolumn{1}{c}{} & \multicolumn{1}{c}{(GHz)} & \multicolumn{1}{c}{(\kms)}& \multicolumn{1}{c}{(mJy)} & \multicolumn{1}{c}{(\kms)}& \multicolumn{1}{c}{(mJy)} \\  \hline
			CH$_3$OH & $17_{-2} \rightarrow 18_{-3}$\,E  & 6.181146(21)$^{[1]}$ & 40 -- 140& 6.7 & --40 -- 50  & 9.1 \\    
			& $5_{1} \rightarrow 6_0$\,A$^{+}$  &  6.6685192(8)$^{[2]}$ & \textbf{40 -- 140}  & \textbf{7.5} & \textbf{--40 -- 60} & \textbf{9.8} \\ 
			& $12_{4} \rightarrow 13_3$\,A$^{-}$  &  7.682246(50)$^{[3]}$  & 24 -- 104 & 4.9&  --40 -- 50  & 5.6\\    
			& $12_{4} \rightarrow 13_3$\,A$^{+}$   &   7.830848(50)$^{[3]}$ & 50 -- 140 & 5.3& --40 -- 40 & 5.8\\
			&$9_{-1} \rightarrow 8_{-2}$\,E  & 9.936202(4)$^{[4]}$ & -   & - & 2 -- 42 & 6.3 \\    
			&  $4_{3} \rightarrow 5_{2}$\,A$^{-}$ &   10.058257(12)$^{[4]}$ & -  & - & 4 -- 40 & 6.0 \\
			&  $2_{1} \rightarrow 3_{0}$\,E &   19.967396(2)$^{[5]}$ & 0 -- 200  & 4.3$^\dagger$ & - & - \\ 
			&  $11_{1} \rightarrow 10_{2}$\,A$^{+}$ &   20.171089(2)$^{[5]}$ & 0 -- 200  & 4.5$^\dagger$ &  -  & - \\ 
			&  $10_{1} \rightarrow 11_{2}$\,A$^{+}$ &   20.970658(37)$^{[6]}$ & 0 -- 200  & 5.5$^\dagger$ &  - & - \\
			&  $12_{2} \rightarrow 11_{1}$\,A$^{-}$ &   21.550342(42)$^{[6]}$ & 0 -- 200  & 6.4$^\dagger$ & - & - \\ 
			&  $9_{2} \rightarrow 10_{1}$\,A$^{+}$ &   23.121024$^{[5]}$ & 0 -- 200  & 7.0$^\dagger$ & - & - \\ 
			&  $10_{1} \rightarrow 9_{2}$\,A$^{-}$ &   23.444776(2)$^{[5]}$ & \textbf{0 -- 200}  & \textbf{6.4$^\dagger$} & - & - \\ 
			H$_2$CO & $1_{1,0} \rightarrow 1_{1,1}$  & 4.829657(2)$^{[7]}$  & \textbf{75 -- 124}  & \textbf{8.7} & --25 -- 55 & 11.9 \\
			H$_2$O & $6_{1,6} \rightarrow 5_{2,3}$  & 22.2350771(1)$^{[8]}$  & \textbf{0 -- 200}  & \textbf{8.2$^\dagger$} & - & - \\
			NH$_3$ & $6_3 \rightarrow 6_3$ & 19.757538(5)$^{[9]}$  & 0 -- 200  & 4.2$^\dagger$ & - & - \\
			& $7_5 \rightarrow 7_5$  & 20.804830(5)$^{[9]}$   & 0 -- 200 & 5.4$^\dagger$ &  - & - \\
			& $10_8 \rightarrow 10_8$ &   20.852527(5)$^{[9]}$ & 0 -- 200  & 5.4$^\dagger$ & - & -\\
			& $11_9 \rightarrow 11_9$ &   21.070739(5)$^{[9]}$ & 0 -- 200  & 5.8$^\dagger$ & - & -\\
			& $6_5 \rightarrow 6_5$  & 22.732429(5)$^{[9]}$   & 0 -- 200 & 7.2$^\dagger$ & -  & - \\
			& $9_8 \rightarrow 9_8$ &   23.657471(5)$^{[9]}$ & 0 -- 200  & 6.2$^\dagger$ & - & -\\ 
			& $3_3 \rightarrow 3_3$  &  23.872453(5)$^{[10]}$   & \textbf{0 -- 200} & \textbf{6.1$^\dagger$} & -  & - \\
			& $4_4 \rightarrow 4_4$ &   24.1394169(1)$^{[11]}$ & 0 -- 200 & 5.7$^\dagger$ & - & -\\ 
			OH & $^{2}\Pi_{1/2},~J=1/2$ & 4.765562(3)$^{[12]}$   & -  & - & --34 -- 46 & 9.5 \\
			& $^{2}\Pi_{3/2},~J=5/2$  & 6.030747(5)$^{[13]}$   & 50 -- 140  & 6.9 & --40 -- 50  & 9.0 \\
			& $^{2}\Pi_{3/2},~J=5/2$ &   6.035092(5)$^{[13]}$ & 40 -- 140  & 6.8 & --20 -- 45 & 9.2 \\ \hline
		\end{tabular} \label{tab:rest_freq}		
	\end{center}
	\begin{flushleft}
		Note: $^{[1]}$\citet{Pickett+98}, $^{[2]}$\citet{Muller+04}, $^{[3]}$\citet{Tsunekawa+95}, $^{[4]}$\citet{Breckenridge+95}, $^{[5]}$\citet{Mehrotra+85}, $^{[6]}$\citet{Xu+97},$^{[7]}$\citet{Kukolich75}, $^{[8]}$\citet{Kukolich+69}, $^{[9]}$\citet{Poynter+75}, $^{[10]}$\citet{Kukolich+67}, $^{11}$\citet{Kukolich+70}, $^{[12]}$\citet{Radford68} and $^{[13]}$\citet{Meerts&75}. \\
	\end{flushleft}	
\end{table*}

The observations reported in this paper were made using the Australia Telescope Compact Array (ATCA). Observations of G\,24.33+0.13 were taken as a part of the C3321 non a-priori assignable (NAPA) project on 2019 November 26/27 and during Director's time on 2021 February 21. The single epoch for G\,359.62--0.24 was observed during Director's time on 2020 July 25/26. All epochs of observation using the 4--10~GHz frequency setup used the CFB\,1M-0.5k mode of the Compact Array Broadband Backend \citep[CABB; ][]{Wilson+11}. This CABB configuration allows for two 2~GHz wide bands ($2048 \times 1$~MHz channels) and up to 16 1~MHz zoom bands (each consisting of $2048\times0.5$~kHz channels) in each wide band. Zoom bands are `stitched' together in order to achieve appropriate velocity coverage where required. The 19 -- 24~GHz frequency setup utilised the CFB\,64M-32k CABB mode, which consists of two 2~GHz ($32 \times 64$~MHz channels) wide bands and up to 16 64~MHz zoom bands (each consisting of $2048\times32$~kHz channels). 

The ATCA was configured in the 1.5C array configuration for the 2019 epoch of G\,24.33+0.13 (minimum and maximum baselines of 77 and 1485 m respectively, with ca06 flagged) and 6D configuration for the 2021 epoch (minimum and maximum baselines of 77 and 5878 m respectively). For G\,359.62--0.24, the array was in the hybrid H214 configuration (minimum and maximum 82 and 247 m respectively, with ca06 flagged). The synthesised beam sizes achieved for each observation epoch are tabulated in Table \ref{tab:array_config}, alongside the phase centre coordinates and total time on-source.

Flux density and bandpass were calibrated with respect to PKS\,B1934--638 and PKS\,B1253--055 respectively for all observation epochs. Unresolved quasar sources for phase calibration were chosen from the ATCA Calibrator Database based on brightness, compactness and proximity to our target sources. TXS\,1829--106 was selected for G\,24.33+0.13 and TXS\,1714--336 for G\,359.62--0.24. All epochs utilised the same observing strategy of starting with a 10 minute integration on the bandpass calibrator, interleaving 10 minute target scans with 1 minute and 40 seconds on the relevant phase calibrator and finally finishing with a 10 minute scan on the flux density calibrator.

Both sources were searched for lines using the ATCA C/X band receiver, while K band observations were only made toward G\,24.33+0.13. Table \ref{tab:rest_freq} contains information on all molecular transitions searched towards G\,24.33+0.13 and G\,359.62--0.24. Director's time observations of G\,359.62--0.24 were made using two different C/X band frequency setups which allowed for the inclusion of some additional C/X lines compared to G\,24.33+0.13. Target positions for the 6.7-GHz class~II methanol masers were taken from the methanol multibeam (MMB) survey \citep{Caswell+10, Breen+15}. For G\,24.33+0.13, poor uv-coverage of our observations (limited hour angle coverage and equatorial declination of G\,24.33+0.13), results in a synthesised beam that is highly elongated in the north-south direction even for the epoch utilising a hybrid array configuration. This results in small uncertainty for the right ascension ($\sim 0.5''$), and comparatively large uncertainty in the declination values ($> 3''$) we report in Table \ref{tab:spectral_line}. The first epoch (2019 November) of G\,24.33+0.13 data was taken approximately 3 months (82 days) after the flaring was first reported toward this source, with our second, follow-up epoch (2021 February) 15 months later. This secondary epoch allows us to investigate the quiescent state of G\,24.33+0.13 and determine how the detected lines evolved after the flaring event subsided.

Data were reduced using {\sc miriad}, following standard techniques for ATCA spectral line observations, with primary, secondary and bandpass calibration utilising the sources outlined above. The {\sc miriad} uvlin task was used to estimate the intensity of line-free spectral channels on each baseline and consequently subtract any contribution from continuum emission away from the spectral line emission. Phase and amplitude self-calibration were performed (where applicable) on any detected emission from the continuum subtracted data, using the brightest components for each pointing. Spectral line data from each transition were imaged over the maximum possible velocity range (dependent on constraints of our frequency setup) with various spectral resolutions while searching for evidence of emission.  For reference purposes, details of the final spectral line cubes (including $1\sigma$ RMS noise values), imaged with a spectral resolution of 0.1~\kms,  can be found in Table \ref{tab:rest_freq}. The {\sc miriad} task imfit was used to extract locations and peak flux density values for the detected emission, imfit reports the peak value and location of a two-dimensional Gaussian fit to a single velocity plane within the continuum subtracted spectral line cube.

\section{Results} \label{sec:results}

\begin{table*} 
	\begin{center}
		\caption{Details of the detected spectral line emission. Velocity is with respect to local standard of rest. Locations and peak flux density of maser emission were extracted using the imfit {\sc miriad} task on the brightest component from the spectral line cubes. Positional uncertainties (from fitting) are $\sim0.5''$, except for declination values for G\,24.33+0.13 which are $\sim3''$. }
		\begin{tabular}{@{}lllrcccccl@{}}
			\toprule
			\multicolumn{1}{c}{\bf Target} &   \multicolumn{1}{c}{\bf Epoch} &  \multicolumn{1}{c}{\bf Molecular} & \multicolumn{1}{c}{\bf Line} & \multicolumn{1}{c}{\bf R.A.}  & \multicolumn{1}{c}{\bf Dec.}  & \multicolumn{2}{c}{\bf Flux Density}  &\multicolumn{1}{c}{\bf Peak}     & \multicolumn{1}{c}{\bf Velocity}  \\
			\multicolumn{1}{c}{\bf Source} &   \multicolumn{1}{c}{}  &   \multicolumn{1}{c}{\bf Species}   & \multicolumn{1}{c}{} &\multicolumn{1}{c}{(J2000)}   & \multicolumn{1}{c}{(J2000)} &  \multicolumn{1}{c}{\bf Peak} & \multicolumn{1}{c}{\bf Integrated}  & \multicolumn{1}{c}{\bf Velocity}&  \multicolumn{1}{c}{\bf Range}  \\
			\multicolumn{1}{c}{} & & &  & \multicolumn{1}{c}{\bf $h$~~~$m$~~~$s$}& \multicolumn{1}{c}{\bf $^\circ$~~~$\prime$~~~$\prime\prime$} & \multicolumn{1}{c}{(Jy)} &\multicolumn{1}{c}{(Jy~\kms)}   & \multicolumn{1}{c}{(\kms)}  &  \multicolumn{1}{c}{(\kms)}   \\ \midrule
			G\,24.33+0.13 & 2019 Nov & CH$_3$OH  &     6.7-GHz   & 18 35 08.1   & $-$07 35 04.5  & 60.1  & 21.9  & 115.3 & 107 -- 121 \\
			 & &  &     23.4-GHz   &  18 35 08.0  & $-$07 34 57.6 & 1.13 & 0.93 & 113.2 & 112.8 -- 113.6 \\
			& & H$_2$CO &    4.8-GHz   & 18 35 08.1  & $-$07 35 05.9  &  0.62  &  0.28 & 109.3  & 107 -- 117  \\
			& & H$_2$O &   22.2-GHz   & 18 35 08.1  & $-$07 35 06.5  & 28.2 & 71.3 & 124.8 & 53 -- 128 \\
			& & NH$_3$ &   23.9-GHz   & 18 35 07.8 & $-$07 35 09.8 & 0.76 &  0.82 & 113.5 & 112.8 -- 114.4 \\
			 & 2021 Feb & CH$_3$OH  &     6.7-GHz   & 18 35 08.1   & $-$07 35 07.0  & 6.23  & 5.7  & 110.2 & 107 -- 121 \\ 
			& & H$_2$CO &    4.8-GHz   & 18 35 08.1  & $-$07 35 05.3  &  0.67 &  0.18 & 109.6  & 108 -- 111  \\ 
			G\,359.62--0.24  &  2020 July & CH$_3$OH & 6.7-GHz  &  17 45 39.1 & $-$29 23 30.5 & 234.2 & 32.9 & 19.5 & 18 -- 25 \\  \hline           
		\end{tabular} \label{tab:spectral_line}
	\end{center}
\end{table*}

\subsection{G\,24.33+0.13}

We detected emission from the 6.7-GHz ($5_{1} \rightarrow 6_0$\,A$^{+}$) and 23.4 ($10_{1} \rightarrow 9_2$\,A$^{-}$) methanol, 22.2-GHz ($6_{1} \rightarrow 5_2$) water, 23.9-GHz ($3_{3} \rightarrow 3_3$) ammonia and 4.8-GHz formaldehyde ($1_{1,0} \rightarrow 1_{1,1}$) transitions toward G\,24.33+0.13. Details of the spectral line emission from these transitions, along with the epoch they were observed in is tabulated in Table \ref{tab:spectral_line}. The position of the methanol maser emission is consistent with previously reported values from the MMB survey \citep{Breen+15}, to within our astrometric accuracy ($\sim0.5''$ and $\sim3''$ in R.A. and declination respectively). A large number of other potential maser transitions were included in observations toward G\,24.33+0.13, but none other than the above resulted in clear detections. Details of the final spectral line cubes for all transitions, including the RMS noise values for a 0.1/0.5~\kms channel (dependent on frequency setup), are listed in Table \ref{tab:rest_freq}. It should be noted that during preparation of this manuscript, private communication revealed that the first detection of formaldehyde maser emission in this source was made by the Tianma 65m Radio Telescope on 2019 September 7 by members of the M2O group (Chen et al. in preparation).

\begin{figure*}
	\centering
	\includegraphics[scale=0.65]{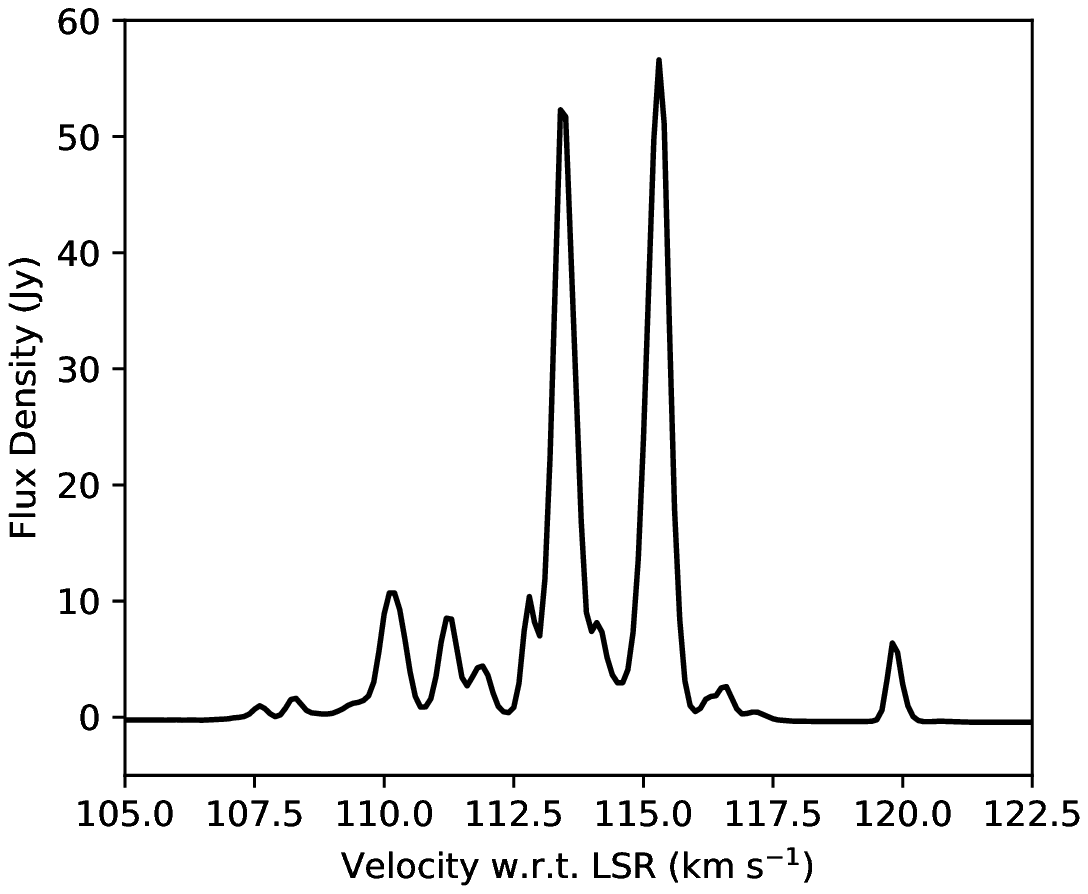}
	\includegraphics[scale=0.65]{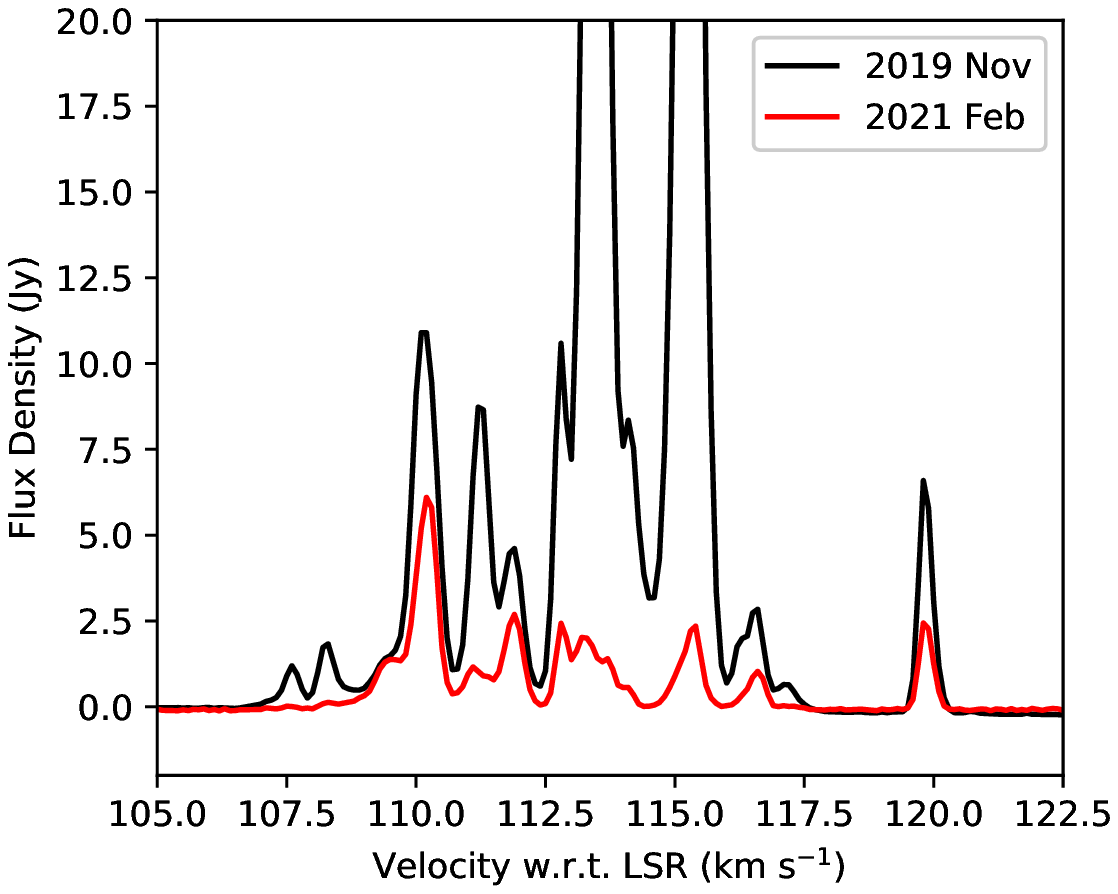}
	\caption{\textit{Left:} Spectrum of 6.7-GHz class~II maser emission toward G\,24.33+0.13 during the 2019 November flaring epoch. The velocity range has been restricted to the range over which we detect signal. \textit{Right:} A comparison between 6.7-GHz class~II maser spectra from our 2019 November and 2021 February epochs. }
	\label{fig:g24_6.7}
\end{figure*}

The 6.7-GHz methanol emission consists of a series of maser components spread over the velocity range 107--121~\kms. During the flare, the spectral profile is dominated by two primary peaks at 113.4 and 115.3 \kms, with several weaker components (factor of $\sim6$ lower flux density) distributed across the rest of the velocity range (see Figure \ref{fig:g24_6.7}). Our second epoch of data, (17 months after the flare was detected), reveals that the flux density of these two peaks has decreased by a factor of $\sim20$, with the 110~\kms\ feature becoming the brightest maser component (see Figure \ref{fig:g24_6.7}). In addition to the two flaring features dimming, we also see a general decrease in flux density across all maser components in the second epoch. Line widths of the maser components are all approximately 0.5 \kms.  

\begin{figure}
	\centering
	\includegraphics[scale=0.65]{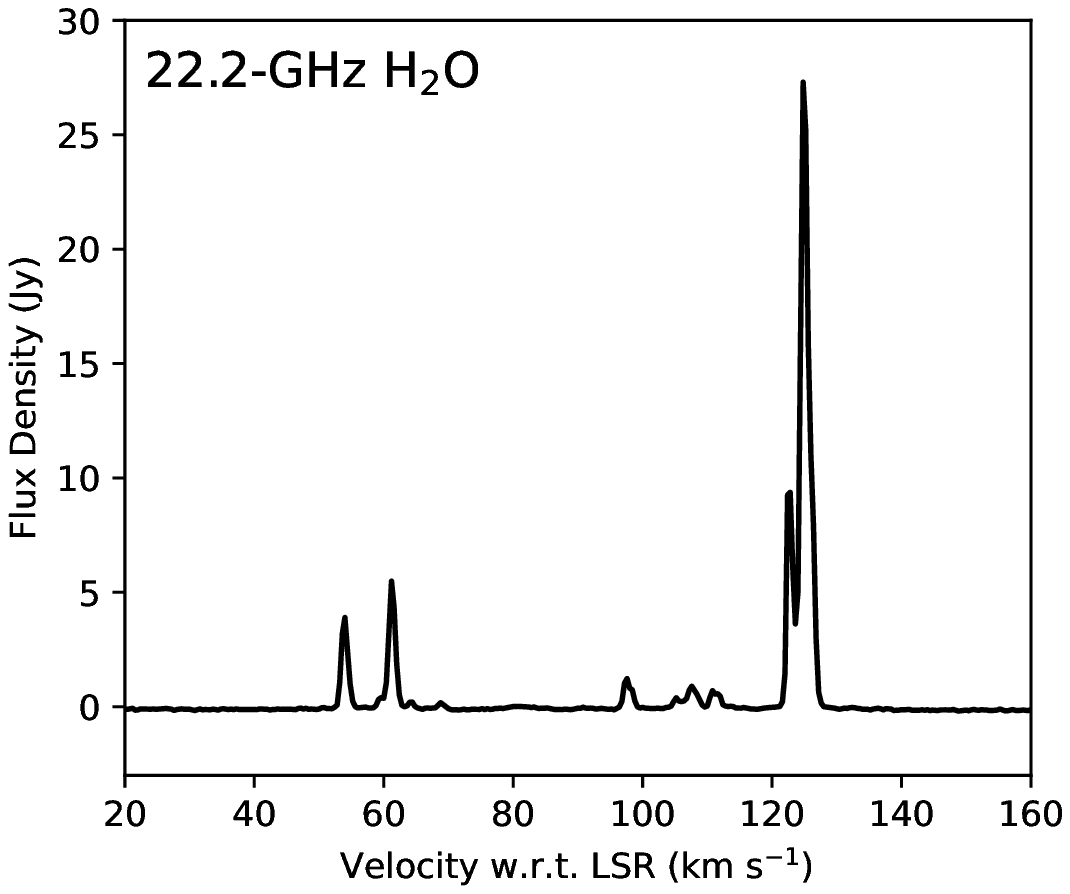}\\
	\includegraphics[scale=0.65]{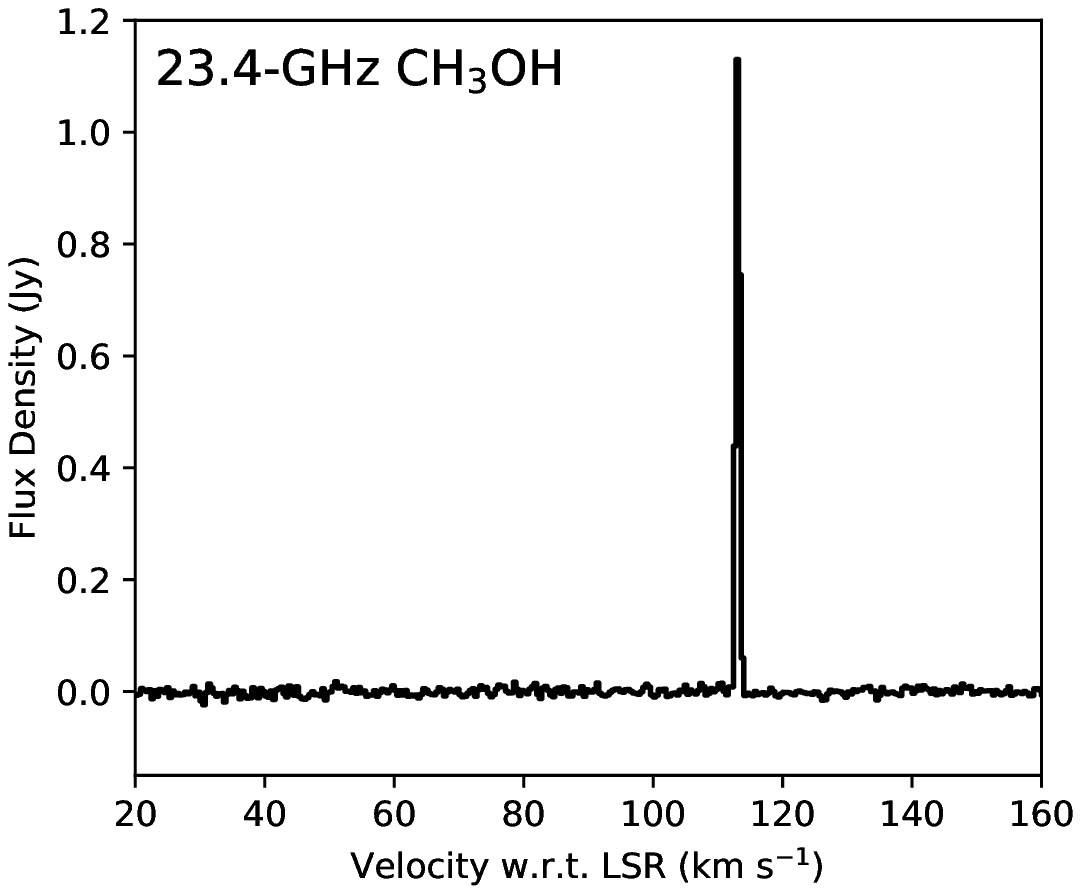}\\
	\includegraphics[scale=0.65]{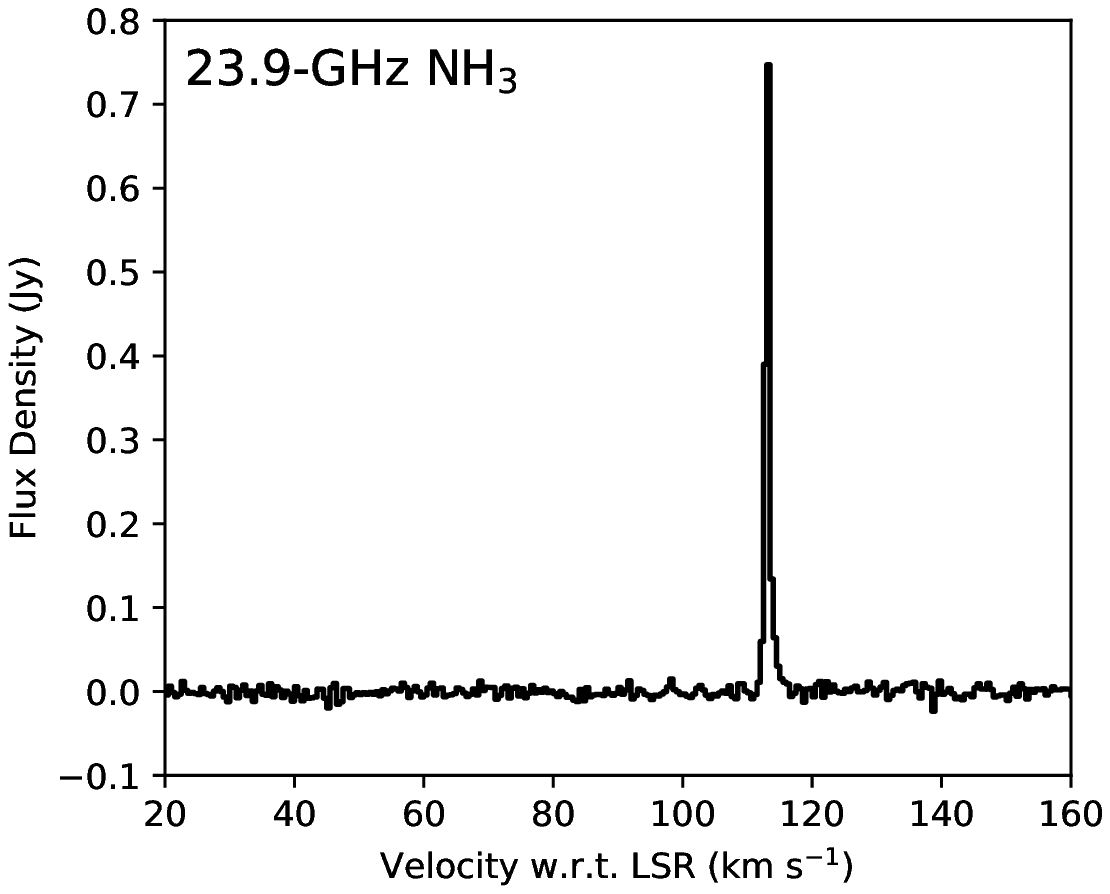}
	\caption{Spectrum of 22.2-GHz water maser emission (top), 23.4-GHz class~I methanol maser emission (middle) and 23.9-GHz class~I ammonia maser emission toward G\,24.33+0.13 during the 2019 November flaring epoch.}
	\label{fig:g24_20ghz_lines}
\end{figure}

We observe the 22.2-GHz water maser emission to be spread over a broad velocity range, with similar coverage (from $\sim50-130$~\kms) to the most recent prior observations reported by \citet{Cyganowski+13}. We observe a spectral profile dominated by a 2 component red-shifted feature, approximately twice the flux-density of any previously reported 22.2-GHz water maser feature from this source (see top panel of Figure \ref{fig:g24_20ghz_lines}). However, the inherent variability of the H$_2$O maser line prevents meaningful comparison between our post-flare observation and those from several years prior to the event.

\begin{figure}
	\centering
	\includegraphics[scale=0.65]{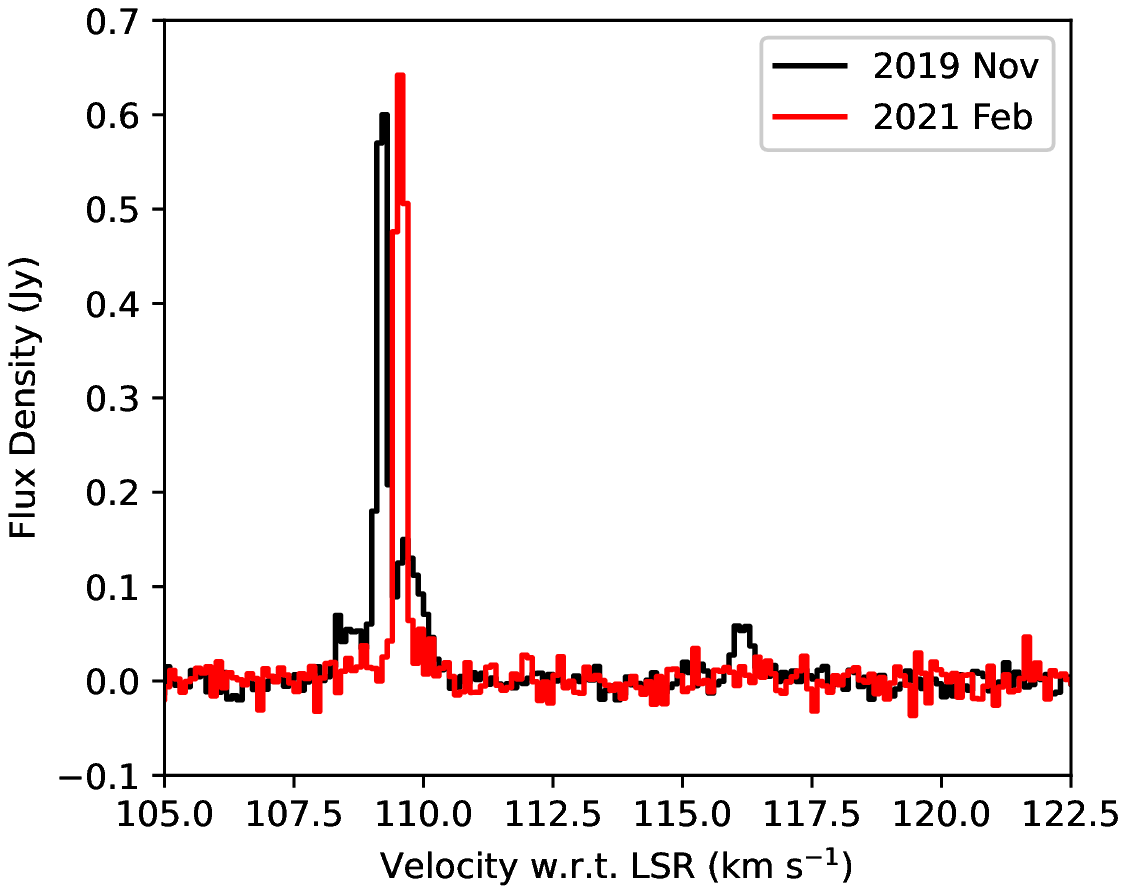}
	\caption{
	Comparison of the 4.8-GHz formaldehyde spectra emission toward G\,24.33+0.13 from the 2019 November (black) and 2021 February (red) observation epochs.
	}
	\label{fig:g24_4.8}
\end{figure}

The position of the peak formaldehyde emission is consistent (to within our astrometric accuracy) between the two epochs, and at approximately the same location within the star-forming region as the class~II methanol maser emission. For the 2019 epoch, the formaldehyde emission spectrum consists of a primary component with a full-width at half maximum of $\sim0.2$~\kms\ at a velocity of 109.3~\kms, along with what appears to be 3 much weaker features (see Figure \ref{fig:g24_4.8}). The follow-up epoch from 2021 shows a simplification of this spectral profile to a single narrow component at approximately the same flux density ($\sim0.6$~Jy), redshifted with respect to the 2019 peak by 0.3 \kms\ (see Figure \ref{fig:g24_4.8}). Based on the narrow line width seen from these spectral profiles it is likely that this emission is the result of a maser process. This interpretation will be discussed further in Section \ref{sec:masing_formaldehyde}.

The spectral profile of the 23.4-GHz methanol transition consists of a single 1.13~Jy component at $113.2$~\kms (see middle panel of Figure \ref{fig:g24_20ghz_lines}), with a FWHM of $\sim0.8$~\kms (this FWHM is an upper limit due to the limited spectral resolution of our frequency setup). The emission is located approximately 7 arcseconds north and 2 arcseconds west of the peak 6.7-GHz emission. As mentioned previously, our observing set up is not particularly sensitive to north-south offsets, however, 7 arcseconds is large enough that we are confident this emission is indeed offset from the class~II methanol maser emission. The 23.4-GHz methanol transition has only been found toward three other sources within the Milky Way, G\,357.97--0.16, G\,343.12--0.06 \citep{Voronkov+11} and G\,305.21$+$0.21 (Voronkov, private communication), where it was determined to be a class~I methanol maser.

Similar to the 23.4-GHz methanol, the ammonia (3,3) spectral profile (bottom panel of Figure \ref{fig:g24_20ghz_lines}) is made up of a single component at $113.5$~\kms, with a flux density of 0.76~Jy and FWHM of $\sim$0.8~\kms (similarly to the 23.4-GHz emission, this is an upper limit). The position of the peak emission is also offset by approximately $5$ arcseconds south-west of the class~II methanol emission.  

\subsection{G\,359.62--0.24}

Maser emission from the 6.7-GHz methanol ($5_{1} \rightarrow 6_0$\,A$^{+}$) transition was detected toward G\,359.62--0.24, with details of this emission tabulated alongside the G\,24.33+0.13 information in Table \ref{tab:spectral_line}. Table \ref{tab:rest_freq} contains the 0.1~\kms\ channel RMS noise values for the spectral line cubes of both the detected and non-detected transitions searched for toward G\,359.62--0.24.

\begin{figure}
	\centering
	\includegraphics[scale=0.65]{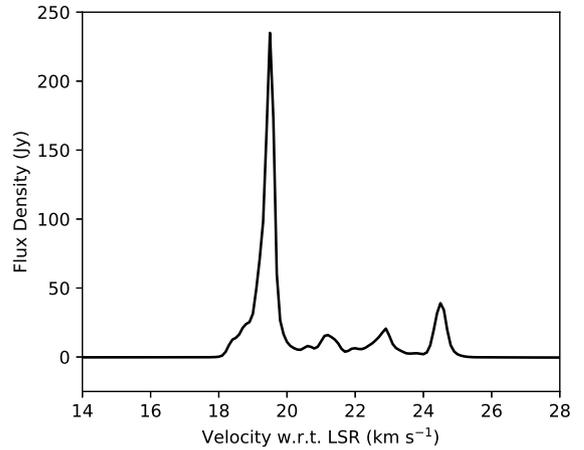}
	\caption{Spectrum of 6.7-GHz class~II maser emission toward G\,359.62--0.24. 
	}
	\label{fig:g359_6.7}
\end{figure}

6.7-GHz class~II maser emission is detected over the velocity range 18 -- 25~\kms, with the primary maser component at 19.5~\kms\ (see Figure \ref{fig:g359_6.7}). The maser components have line widths of approximately 0.5 \kms. This class~II emission is at the location of the phase centre for our observations and the reported 6.7-GHz maser position from the MMB survey \citep{Caswell+10}. 

\section{Discussion}

\subsection{Comparison with previous class~II maser observations toward these sources}

\subsubsection{G\,24.33+0.13}

Comparing our flaring 6.7-GHz spectrum to that of the 2010 December flaring event reported by \citet{Wolak+18}, we see that the spectral profile is very similar despite the two epochs being almost a decade apart. During this previous flare in G\,24.33+0.13, flaring was most prominently seen from the same two features that dominate the spectral profile of our observations (line velocities of 113.4 and 115.3\,\kms), with the flux density of these components increasing by over an order of magnitude compared to quiescent phases \citep{Breen+15, Yang+19}. Additionally, the peak flux density that the features reach are very similar between the two flaring events. Minor variations are also observed from the other maser components, though notably, the flux density of the 110 and 112~\kms\ features appear to be relatively consistent with values reported prior to flaring. We also see the existence of the maser components at line velocities of approximately 108  and 116.5~\kms, which were only detectable during the peak of the previous flare \citep{Wolak+18}.

G\,24.33+0.13 was observed as part of the MMB survey in 2009 \citep{Breen+15}. The spectral profile of the MMB observations is consistent with that reported by \citet{Wolak+18} before the 2010 flaring event, with the 110~\kms\ component significantly brighter (factor of $\sim2$) than any other emission components. \citet{Yang+19} present Tianma observations of the 6.7-GHz class~II methanol emission toward G\,24.33+0.13 from 2016 August. Again, the spectral profile is similar to previous quiescent phase observations, however, in this case the relative flux density between the 110~\kms\ component and the other maser components is more pronounced (factor of $\sim3$ brighter). Comparing our two primary components to those with the same line velocity from \citet{Yang+19} we see an increase in flux density by more than a factor of 25. Comparing the 6.7-GHz class~II spectrum from our post flare epoch, we see the methanol emission returning to the same spectral profile that has been previously observed during quiescent phases \citep{Breen+15, Yang+19}, with the flux density of all features decreasing. Again, the 110~\kms\ feature appears to be the least affected by the variability, however, it still shows a flux density decrease of $\sim40\%$ between the 2019 November and 2021 February epochs we report here.

\subsubsection{G\,359.62--0.24}

Variability in G\,359.62--0.24 was originally reported by \citet{Caswell+95a} with subsequent long term monitoring of the 6.7-GHz class~II emission toward G\,359.62--0.24 carried out by \citet{Goedhart+04} over the period from 1999 January to 2003 March. Their monitoring reveals a significant decrease in the flux density of all maser components during this time. In addition to the general decrease in maser flux density during this period, a significant flare ($\sim3$ to $\sim11$ Jy) from the 18.4~\kms\ maser component was observed during 2000 January, lasting approximately 500 days. Even when imaging with the finest spectral resolution possible from our observations ($\sim 0.03$~\kms), we can not easily spectrally resolve the 18.4~\kms\ component in our observations, however, the spectral profile between 18 and 20~\kms\ does appear to consist of 3 or more maser components (see Figure \ref{fig:g359_6.7}). If this is the case, the 18.4~\kms\ component would be at approximately the same intensity (11--12 Jy) as seen during the flare reported by \citet{Goedhart+04}.

The 6.7-GHz emission toward G\,359.62--0.24 has been observed as part of the MMB survey \citep{Green+09a, Caswell+10}. Comparison of their spectrum to ours reveals a very similar spectral profile, however we see flux density increases in the 19.5 and 24.5\,\kms\ components (by a factor of $\sim6$ and $\sim2$, respectively). The MMB spectrum is from 2008 March, over a decade before the observations we report here, therefore, it is not possible for us to determine exactly what changes in the spectral profile result from the flare event, and what are due to intrinsic variability of the source. However, the peak flux density of 234~Jy from the 19.5~\kms component is over a factor of 5 higher than has ever been reported from this source \citep{Goedhart+04, Caswell+10}, and so it is likely that the flux density of this component has increased during this flare event.

\subsection{The fourth detection of 23.4-GHz class~I methanol maser emission in the Milky Way} \label{sec:masing_classI}

Maser emission in the 23.4-GHz class~I methanol maser line has been found 
 toward three other sources to date \citep[G\,357.97--0.16, G\,343.12--0.06 and G\,305.21+0.21; ][Voronkov private communication]{Voronkov+11}. \citet{Voronkov+11} determined this transition to be class~I, based on pumping models from \citet{Cragg+92}, close positional alignment  with other known class~I masers and the detection of the line in absorption toward W3(OH) by \citet{Menten+85}. Class~I methanol masers tend to be offset from the radiatively pumped class~II methanol masers (and therefore, the YSOs within the HMSF regions), and associated with shocks from outflows or cloud-cloud collisions within their host sources \citep{Cyganowski+09, Cyganowski+12, Sjouwerman+10, Voronkov+14, McCarthy+18}. The offset we observe between the 23.4-GHz emission and the 6.7-GHz class~II maser emission from G\,24.33+0.13 is consistent with this expectation, however, as mentioned previously, our array-configuration is not sensitive to north-south offsets and follow-up observations are required to better map the maser components in this source. G\,24.33+0.13 is also host to an extended green object (EGO), categorised by \citet{Cyganowski+08} as a region of excess in the \textit{Spitzer} 4.5~$\mu m$ band that is indicative of outflowing from the YSOs \citep{Cyganowski+09}. A map of this EGO is presented in Figure 10 of \citet{Lee+13}, which shows an enhancement of the EGO north-west of the \ionhy region, at the same location from which we detect the 23.4-GHz methanol emission.

The upper limit we place on the FWHM for the 23.4-GHz methanol component ($\sim0.8$ \kms) is also a factor of 5 narrower than any thermal emission that has been previously observed toward G\,24.33+0.13 \citep{Rathborne+11}. This narrow linewidth, combined with probable association with an outflow and the fact that 23.4-GHz methanol has never been detected as thermal emission indicate that this emission is likely a maser.

\subsection{Masing formaldehyde toward G\,24.33+0.13} \label{sec:masing_formaldehyde}

Prior to this work, only ten Galactic H$_2$CO masers had been detected \citep[e.g. ][]{Araya+06, Chen+17}. In this section we will justify our conclusion that the H$_2$CO emission reported here is the result of a maser process.

The spectral profile of the emission in the 2019 November observations seems to consist of 4 narrow emission components, with the FWHM of the primary component being approximately 0.2 \kms. This narrow line width is consistent with the other known examples of masing formaldehyde \citep[e.g. ][]{Araya+06,Chen+17}. While absorption is commonly observed from the 4.8-GHz formaldehyde transition, thermal emission from the 4.8-GHz H$_2$CO line has only been reported from one source, Orion KL \citep{Zuckerman+75, Mangum+93}. The reported line width of 3~\kms\ for this thermal emission is more than an order of magnitude greater than what we observe toward G\,24.33+0.13 \citep{Zuckerman+75}. Additionally, when comparing our two epochs, we see the maser changing from multiple emission components to a single bright component of emission over the 15 month span. Variability like this is not observed from thermal lines, and is a strong argument that the formaldehyde is masing.

 The pumping mechanism responsible for population inversion in 4.8-GHz H$_2$CO masers is still poorly understood, with both collisional and radiative origins proposed \citep{Hoffman+03, Araya+05, Araya+06}. Correlated variability between H$_2$CO and class~II methanol masers has suggested that these masers may share similar radiative excitation mechanisms \citep{Araya+10}, and so far all H$_2$CO masers have been observed towards high-mass star formation regions that also host class~II methanol masers. While not always spatially coincident, masers from these two species tend to be distributed over similar distances from the excitation sources \citep{Araya+10, Chen+17}. Toward G\,24.33+0.13, the individual components of the H$_2$CO share line velocities with the detected 6.7-GHz class~II methanol maser components, indicating they are likely distributed over a similar region in this source. However, VLBI observations of the H$_2$CO emission would be required to confirm this.

European VLBI network (EVN) observations of this source by \citet{Bartkiewicz+16}, determine a total angular extent of approximately $0.5$~arcseconds for the 6.7-GHz class~II methanol maser spots. Assuming, as stated above, that the H$_2$CO is distributed over a similar region, this results in a lower limit on the brightness temperature for the peak H$_2$CO emission of more than $10^5$~K. As the H$_2$CO is unresolved on our longest baseline (5878 m), we can consider a more conservative angular scale of $2''$ for which we calculate a brightness temperature larger than $10^4$~K. Even the conservative estimate of the brightness temperature implies that this emission is the result of a maser process.

Our two epochs do not allow us to determine whether this masing formaldehyde showed evidence of flaring alongside the class~II methanol maser emission. The integrated flux density decreases by $\sim35$\% in the 15 months between our observations, which is far less dramatic than the $\sim75$\% seen from the methanol. Chen et al. (in preparation) will report monitoring of this 4.8-GHz maser with higher temporal resolution across the class~II methanol maser flare period in a future publication.

\subsection{Ammonia (3,3) emission toward G\,24.33+0.13} \label{sec:masing_ammonia}

The metastable ($J = K$) ammonia (3,3) inversion transition has been readily observed as thermal emission \citep[e.g. ][]{Wienen+12, Cyganowski+13} and much more rarely (8 sources total) as maser emission towards star-formation regions in the Milky Way \citep{Mangum&94, Kraemer&95, Zhang&95, Zhang+99, Hunter+08, Brogan+11, Urquhart+11, Walsh+11}. Masers from metastable ammonia transitions rely on collisional pumping in order to achieve population inversion. \citet{Cyganowski+13} reported emission from the (1,1), (2,2) and (3,3) ammonia transitions toward G\,24.33+0.13, with an LSR velocity of 113.7~\kms. Our interferometric observations detect the (3,3) line at approximately the same LSR velocity, though our peak flux-density is $\sim30\%$ lower and the upper-limit linewidth for the component is much narrower than the previously observed (3,3) emission. While this narrower linewidth and offset location hint at a potential maser origin, analysis of the emission from each baseline reveals a clear relationship between the spectral profile and baseline length, with our shortest baseline (77 metres) spectrum being comparable in both linewidth and flux-density to that reported by \citet{Cyganowski+13}, and the emission completely resolved out on our 1000+ metre baselines. This relationship may be the result of the ammonia (3,3) emission consisting of both a compact (narrow linewidth) and extended (broad linewidth) component, however, further modelling and observations would be required to conclusively determine this.

\subsection{Flare mechanisms and comparison to the accretion burst in G\,358.93--0.03}

The recent accretion burst related 6.7-GHz class~II methanol flare in G\,358.93--0.03 was accompanied by dramatic flaring in various other class~II methanol maser transitions, several of which had never been detected toward an astronomical source previously \citep[e.g.][]{Breen+19}. Similar flaring of rare maser lines (methanol and OH) has also been observed alongside the accretion burst in NGC~6334I \citep{Macleod+18}. We included a handful of these class~II methanol lines in our observations (see Table \ref{tab:rest_freq}), however, did not see emission from them toward either flaring source. Based on the investigation of the G\,358.93--0.03 flare event, emission from these rare methanol transitions (6.2-, 7.7- and 7.8-GHz ) had a decay half-life of approximately 10 days \citep{Chen+20}. Assuming similar relative flux density values between the 6.7-GHz and these rare transitions as seen toward G\,358.93--0.03 \citep{Breen+19}, we would have been able to easily detect emission from these lines ($>50\sigma$ detections) toward both these sources if it were present.

While we did not detect any of these additional class~II methanol maser transitions, we did observe rare collisionally and radiatively pumped masers toward G\,24.33+0.13. The lack of any contemporaneous flaring of other maser lines toward G\,359.62--0.24 may suggest a different mechanism (rather than a large accretion burst event) may be driving the flaring in this source. However, due to the small sample of maser flares where such an extensive molecular line search has been conducted, this behaviour (flaring observed from a plethora of transitions) may not be representative of a typical accretion-burst flare and our single observation epoch toward G\,359.62--0.24 is not sufficient for us to properly determine the mechanism behind the flaring.

For G\,24.33+0.13, the decrease in the flux density of all maser components (see Figure \ref{fig:g24_6.7}) from our second epoch suggests that the flaring does not result from an increase in the maser path length, where we would expect to see variation in a small number of velocity-coherent maser features \citep[e.g. ][]{Burns+20b}. The similarity between this flare and that observed in 2010 by \citet{Wolak+18} indicates some long-term periodic mechanism may be responsible for flaring in this source. There have been several periodic 6.7-GHz class~II methanol maser sources where the class~II methanol maser flux density is correlated with periodic variation in the infrared emission \citep{Stecklum+18, Olech+19}. In these cases, the infrared radiation field has been identified as scaling with periodic enhancements of the accretion rates in the high-mass YSO, driven by binary interaction. The flaring of G\,24.33+0.13 may be driven by a similar interaction, however, the $\sim9$ year gap between the two reported flare events is a much larger than the 53 to 540 day periods seen toward other periodic class~II methanol maser sources \citep{Olech+19}. 

Both G\,24.33+0.13 and G\,359.62--0.24 have been targets for the iMet maser monitoring program \citep{Yonekura+16}, which has monitored the 6.7-GHz class~II maser emission since 2013. An upcoming publication from this program will produce flare profiles for both of these sources (Yonekura et al. in preparation), allowing for comparison against infrared light curves and further investigation of the flare mechanisms.

\section{Conclusions}

We present the results of an ATCA molecular line search toward two flaring class~II methanol maser sources G\,24.33+0.13 and G\,359.6-0.24. 

Toward G\,24.33+0.13 we observe a factor of 25 increase in the flux density of flaring 6.7-GHz class~II methanol maser components compared with pre- and post-flare observations. A comparison of the 6.7-GHz spectral profile with the previously observed flare from 2010, reported by \citet{Wolak+18}, reveals the same maser components are flaring in both events. Our quiescent phase epoch (15 months later) reveals an across the board decrease in the flux density of the 6.7-GHz class~II methanol masers.

We detect 4.8-GHz formaldehyde emission toward G\,24.33+0.13. The line-width, estimated brightness temperature, similarity with the class~II methanol maser emission and variability in the spectral profile between epochs all indicate that this emission is the result of maser processes. This is the eleventh example of a formaldehyde maser within our Galaxy. Additionally, we report the detection of the fourth example of a 23.4-GHz class~I methanol maser, and detection of ammonia (3,3) emission along with presenting a spectrum of the 22.2-GHz water maser emission shortly after the flare event.


Our observations of G\,359.6-0.24 show no evidence of any contemporaneous flaring of other methanol (or other species) maser transitions. While this is not conclusive, it is inconsistent with other known examples of accretion burst flares, suggesting that the flaring from this source may instead be caused by another mechanism.

\section*{Acknowledgements}

 We thank the anonymous referee for their useful suggestions and comments which helped improve this manuscript. The ATCA is part of the Australia Telescope which is funded by the Commonwealth of Australia for operation as a National Facility managed by CSIRO. This research has made use of NASA's Astrophysics Data System Abstract Service. This research made use of Astropy, a community-developed core Python package for Astronomy \citep{astropy+13}. 
 
 R.A.B. acknowledges support through the EACOA Fellowship from the East Asian Core Observatories Association. 
 
 M.O. thanks the Ministry of Education and Science of the Republic of Poland for granting funds for the Polish contribution to the International LOFAR Telescope (MSHE decision no. DIR/WK/2016/2017/05-1) and for maintenance of the LOFAR PL-612 Baldy (MSHE decision no. 59/E-383/SPUB/SP/2019.1)

\section*{Data availability}

The data underlying this article will be shared on reasonable request to the corresponding author. Australia Telescope Compact Array data is open access 18 months after the date of observation and can be accessed using the Australia Telescope Online Archive (\url{https://atoa.atnf.csiro.au}).

\bibliography{references}

\onecolumn

\appendix

\end{document}